\documentstyle[emulateapj,psfig]{article}

\begin{document}

\title{Relativistic Wind Bubbles and Afterglow Signatures}
\author{Z. G. Dai}
\affil{Department of Astronomy, Nanjing University, Nanjing
210093, China; dzg@nju.edu.cn}

\begin{abstract}
Highly magnetized, rapidly rotating compact objects are widely
argued as central energy sources of $\gamma$-ray bursts (GRBs).
After the GRB, such a magnetar-like object may directly lose its
rotational energy through some magnetically-driven processes,
which produce an ultrarelativistic wind dominated possibly by the
energy flux of electron-positron pairs. The interaction of such a
wind with an outward-expanding fireball leads to a relativistic
wind bubble, being regarded as a relativistic version of the
well-studied Crab Nebula. We here explore the dynamics of this
wind bubble and its emission signatures. We find that when the
injection energy significantly exceeds the initial energy of the
fireball, the bulk Lorentz factor of the wind bubble decays more
slowly than before, and more importantly, the reverse-shock
emission could dominate the afterglow emission, which yields a
bump in afterglow light curves. In addition, high polarization of
the bump emission would be expected if a toroidal magnetic field
in the shocked wind dominates over the random component.

\end{abstract}


\section {Introduction}
The recent observation of high linear polarization during the
prompt $\gamma$-ray emission of GRB 021206 (Coburn \& Boggs 2003)
suggests that GRBs be driven by highly magnetized, rapidly
rotating compact objects. Two popular scenarios for their birth
are the merger of a compact binary or the collapse of a massive
star (for a recent review see M\'esz\'aros 2002). In both
scenarios, a rapidly rotating black hole surrounded by an
accretion disk seems to be a common remnant (Narayan, Paczy\'nski
\& Piran 1992; Woosley 1993; M\'esz\'aros \& Rees 1997a;
Paczy\'nski 1998). However, a millisecond magnetar has also been
argued as an alternative interesting product (Usov 1992; Duncan \&
Thompson 1992; Klu\'zniak \& Ruderman 1998; Dai \& Lu 1998; Spruit
1999; Ruderman, Tao \& Klu\'zniak 2000; Wheeler et al. 2000). To
explain the complex temporal feature, the burst itself, in some of
these energy models, is understood to arise from a series of
explosive reconnection events in a rising, amplified magnetic
field because of the Parker instability. This in fact dissipates
the differentially rotational energy and magnetic energy of the
newborn magnetar or accretion disk.

After the GRB, the remaining object is reasonably assumed to be a
millisecond magnetar or a rapidly rotating black hole surrounded
by an accretion disk. For the latter object, the magnetic field in
the disk could have been amplified initially by differential
rotation to a magnetar-like strength of $\sim 10^{15}$ G, and
particularly, within the framework of the collapsar/hypernova
model, such a field could be kept, due to longevity (with days or
longer) of the disk maintained by fallback of the ejecta. During
the afterglow, the object at the center will directly lose its
rotational energy by the magnetic dipole radiation or the
Blandford-Znajek mechanism.

An energy outflow driven magnetically includes three components:
low-frequency electromagnetic waves, a relativistic wind, and a
toroidal magnetic field associated with the wind. The wind energy
flux is unlikely to be baryon-dominated, because the initial
explosion should have left a clean passage with very few baryon
contamination for a subsequent outflow. The interaction of this
outflow with an outward-expanding fireball implies a continuous
injection of the stellar rotational energy into the fireball. Dai
\& Lu (1998, 2000), Zhang \& M\'esz\'aros (2001) and Chang, Lee \&
Yi (2002) discussed the evolution of a relativistic fireball by
assuming a pure electromagnetic-wave energy outflow, while Rees \&
M\'esz\'aros (1998), Sari \& M\'esz\'aros (2000), Zhang \&
M\'esz\'aros (2002), and Granot, Nakar \& Piran (2003) took into
account a variable and baryon-dominated injection.

However, based on the successful models of the well-observed Crab
Nebula (Rees \& Gunn 1974; Kennel \& Coroniti 1984; Begelman \& Li
1992; Chevalier 2000), a realistic, continuous outflow during the
afterglow is expected to be ultra-relativistic and dominated by
the energy flux of electron-positron pairs. As in the Crab Nebula,
even if an outflow from the pulsar is Poynting-flux-dominated at
small radii, the fluctuating component of the magnetic field in
this outflow can be dissipated by magnetic reconnection and used
to accelerate the outflow, which is eventually dominated by the
energy flux of e$^+$e$^-$ pairs within a larger radius $\sim
10^{17}$ cm (Coroniti 1990; Michel 1994; Kirk \& Skj{\ae}raasan
2003). In the case of an afterglow, therefore, it is natural to
expect that the central object still produces an
ultra-relativistic e$^+$e$^-$-pair wind, whose interaction with
the fireball leads to a relativistic wind bubble. This can be
regarded as {\em a relativistic version of the Crab Nebula}.

In this paper, we explore the dynamics of such a wind bubble and
its emission signatures. In \S 2 we present expressions of the
luminosity of a relativistic wind from a highly magnetized,
rapidly rotating object. In \S\S 3 and 4 we discuss evolution of
the wind bubble and temporal features of the radiation
respectively. In the final section we summarize our findings and
give a brief discussion on implications of the bubble.

\section{The Luminosity of a Relativistic Wind}

We assume that a burst itself arises from a series of explosive
reconnection events. After the GRB, we are left with a highly
magnetized, rapidly rotating compact object. Let's first assume
that it is a millisecond magnetar with period $P$, surface
magnetic field strength $B_s$, moment of inertia $I$, radius
$R_{\rm M}$, and angle between the rotation axis and magnetic
dipole moment $\theta$. Since such a pulsar loses its rotational
energy through the magnetic dipole torque, the luminosity of a
resulting relativistic wind is given by
\begin{equation}
L_w \simeq 4\times 10^{47}B_{\perp,14}^2R_{\rm M,6}^6P_{\rm
ms}^{-4}\,{\rm erg}\,{\rm s}^{-1},
\end{equation}
where $B_{\perp,14}=B_s\sin\theta/10^{14}{\rm G}$, $R_{\rm
M,6}=R_{\rm M}/10^6{\rm cm}$, and $P_{\rm ms}=P/1\,{\rm ms}$.
Because of spin-down, this luminosity will evolve with time as
\begin{equation}
L_w\propto (1+t/T_{{\rm M},0})^{-2}\left \{
       \begin{array}{ll}
         \sim {\rm const.}, & {\rm if}\,\, t<T_{{\rm M},0},\\
         \propto t^{-2}, & {\rm if}\,\, t>T_{{\rm M},0},
        \end{array}
       \right.
\end{equation}
where $t$ is the observer time in units of day, $T_{{\rm
M},0}=0.58B_{\perp,14}^{-2}I_{45}R_{{\rm M},6}^{-6}P_{0,{\rm
ms}}^2$ days is the ``initial" spin-down timescale of the magnetar
at the onset of the afterglow, $P_0=P_{0,{\rm ms}}\times 1\,{\rm
ms}$ is the rotation period at this time, and
$I_{45}=I/10^{45}{\rm g}\,{\rm cm}^2$. Usov (1992) also assumed
that the early spin-down could be due to gravitational wave
radiation besides magnetic dipole radiation. However, the
luminosity for gravitational wave radiation depends on the stellar
ellipticity which is poorly known, and so we neglected the effect
of this mechanism on spin-down.

We now discuss another case in which the central object is a
rapidly rotating black hole surrounded by an accretion disk. If
the amplified magnetic field in the disk does not evolve
significantly with time during fallback of the ejecta, the
rotational energy of this black hole will be gradually extracted
by the Blandford-Znajek mechanism, whose luminosity is
approximated by
\begin{eqnarray}
L_w & = & 1.5\times 10^{51}B_{{\rm BH},15}^2(M_{\rm
BH}/3M_\odot)^2a^2f(a)\,{\rm erg}\,{\rm s}^{-1}\nonumber \\ &
\simeq & 3\times 10^{47}B_{{\rm BH},15}^2(M_{\rm
BH}/3M_\odot)^2(a/0.2)^4\,{\rm erg}\,{\rm s}^{-1},
\end{eqnarray}
where $f(a)=1-[(1+\sqrt{1-a^2})/2]^{1/2}\simeq a^2/8$ for the
rotation parameter $a\ll 1$, $M_{\rm BH}$ is the black-hole mass,
and $B_{{\rm BH},15}$ is the disk field strength in units of
$10^{15}$ G (Lee, Wijers \& Brown 2000). We note that a typical
value ($\sim 10^{47}\,{\rm erg}\,{\rm s}^{-1}$) of the wind
luminosity in equations (1) and (3) has been invoked by Rees \&
M\'esz\'aros (2000) to explain the observed iron lines from some
GRBs within the framework of the collapsar/hypernova model.
Considering the rotational energy of the black hole $E_{\rm
BH}=f(a)M_{\rm BH}c^2\simeq (a^2/8)M_{\rm BH}c^2$ for $a\ll 1$
(Lee et al. 2000), and assuming $\dot{E}_{\rm BH}=L_w$, we find
that the Blandford-Znajek mechanism yields spin-down, similarly to
the magnetar case, if the accreted angular momentum is neglected
because its rate seems to be below the torque driven by the
Blandford-Znajek mechanism at time of days after the burst for
typical parameters in the collapsar/hypernova model. Thus, we
obtain a crude evolution law of the above luminosity,
\begin{equation}
L_w\propto (1+t/T_{{\rm BH},0})^{-2}\left \{
       \begin{array}{ll}
         \sim {\rm const.}, & {\rm if}\,\, t<T_{{\rm BH},0},\\
         \propto t^{-2}, & {\rm if}\,\, t>T_{{\rm BH},0},
        \end{array}
       \right.
\end{equation}
where $T_{{\rm BH},0}=1.0B_{{\rm BH},15}^{-2}(M_{\rm
BH}/3M_\odot)^{-1}(a_0/0.2)^{-2}$ days is the ``initial" spin-down
time of the black hole and $a_0$ is the ``initial" rotation
parameter. One can easily see that equations (2) and (4) are
similar, which implies that our discussion of a relativistic wind
bubble in the remaining text for magnetars should be valid for
black holes.

\section{The Bubble Dynamics in the Thin-Shell Approximation}

Similarly to the Crab Nebula, a rotating magnetar at the center of
an afterglow generates a highly relativistic wind dominated by the
energy flux of e$^+$e$^-$ pairs, with bulk Lorentz factor of
$\gamma_w\sim 10^4-10^7$. Atoyan (1999) argued that the Crab
pulsar initially had $\gamma_w\sim 10^4$ to interpret the measured
radio spectrum of the Crab Nebula. We adopt $\gamma_w=10^4$ as a
fiducial value in our calculations. Because $\gamma_w$ is much
larger than the Lorentz factor of the medium swept up by the
fireball, this wind passes through a shock front and decelerates
to match the expansion velocity of the swept-up medium. Therefore,
a relativistic wind bubble, as a result of interaction of the wind
with the medium, should include two shocks: a reverse shock that
propagates into the cold wind and a forward shock that propagates
into the ambient medium. Thus, there are four regions separated in
the bubble by these shocks: (1) the unshocked medium, (2) the
forward-shocked medium, (3) the reverse-shocked wind gas, and (4)
the unshocked cold wind, where regions 2 and 3 are separated by a
contact discontinuity. For simplicity, we here assume that two
initially-forming forward shocks during interactions of the
fireball both with the medium and with the wind have eventually
merged to one forward shock, and also neglect effects of the
baryon loading whose mass is much less than the swept-up mass,
when the observer time far exceeds the initial deceleration
timescale.

We denote $n_i$ and $P'_i$ as the baryon number density and
pressure of region ``$i$" in its own rest frame respectively, and
$\gamma_i$ is the Lorentz factor of region ``$i$" measured in the
local medium's rest frame. We derive the relative Lorentz factor
of region 3 measured in the rest frame of region 4 as
$\gamma_{34}\simeq
(1/2)(\gamma_w/\gamma_3+\gamma_3/\gamma_w)\simeq
\gamma_w/(2\gamma_3)\gg 1$ for $\gamma_w\gg \gamma_3\gg 1$,
implying a relativistic reverse shock. Because the electron number
density in the rest frame comoving with the unshocked wind is
$n_4=L_w/(4\pi r^2\gamma_w^2m_ec^3)$ (where $r$ is the radius of
region 3 and $m_e$ is the electron mass), according to the jump
conditions for a relativistic shock (Blandford \& McKee 1976), the
pressure of region 3 is calculated by
\begin{equation}
P'_3=\frac{4}{3}\gamma_{34}^2n_4m_ec^2\simeq \frac{L_w}{12\pi
r^2\gamma_3^2c}.
\end{equation}

Neglecting the presence of the reverse shock and the radiative
energy loss of region 2, and assuming an ambient interstellar
medium with constant density of $n_1$, the properties of the
shocked medium in region 2 should satisfy the Blandford-McKee
adiabatic self-similarity solution with the similarity variable at
any radius $r$,
\begin{equation}
\chi=(1+16\gamma_2^2)\left(1-\frac{r}{ct_l}\right),
\end{equation}
where $\gamma_2\equiv \gamma_2(R)$ is the Lorentz factor of the
shocked medium just behind the forward shock whose radius is
denoted as $R$, and $t_l=(R/c)[1+1/(16\gamma_2^2)]$ is the time
measured in the local medium's rest frame. The radius $r$ can thus
be expressed as function of $\chi$ by
\begin{equation}
r=R\left(1+\frac{1}{16\gamma_2^2}\right)\left(1-\frac{\chi}
{1+16\gamma_2^2}\right),
\end{equation}
which implies $r\simeq R$ as long as $\chi\ll 16\gamma_2^2$ for an
ultrarelativistic forward shock. This justifies the {\em
thin-shell approximation}, in which the width of region 2 is
insignificant as compared to the shock radius $R$.

According to Blandford \& McKee (1976), the pressure and Lorentz
factor of the shocked medium at radius $r$ are given by
\begin{eqnarray}
P'_2(r)& = & {4\over 3}n_1m_pc^2\gamma_2^2\chi^{-17/12},\\
\gamma_2(r) & = & \gamma_2\chi^{-1/2},
\end{eqnarray}
where $m_p$ is the proton mass. Along the contact discontinuity,
$\gamma_3=\gamma_2(r)$ and $P'_3=P'_2(r)$, which yield
\begin{equation}
\gamma_3=\gamma_2\chi^{-1/2},
\end{equation}
\begin{equation}
\chi^{-17/12}=\frac{L_w}{16\pi n_1m_pc^3\gamma_2^2\gamma_3^2R^2},
\end{equation}
where the thin-shell approximation $r\simeq R$ has been
considered. For an ultrarelativistic, adiabatic forward shock,
Blandford \& McKee (1976) found its total energy,
\begin{equation}
E_0=\frac{16\pi n_1m_pc^2\gamma_2^2R^3}{17}.
\end{equation}
In deriving the temporal laws of the similarity variable at the
location of the contact discontinuity and the Lorentz factors of
regions 2 and 3, we should note one crucial effect that the
photons that are radiated from regions 2 and 3 at the same time in
the local medium's rest frame will be detected at different
observer times. This is because the Lorentz factor of region 3 is
smaller than $\gamma_2(R)$ by a factor of $\chi^{-1/2}$ so that
for a same time interval in the local medium's rest frame, $R/c$,
the emission from region 3 will reach the observer at time,
\begin{equation}
t\simeq \frac{R}{4\gamma_3^2c},
\end{equation}
and the emission from region 2 will reach the observer at time,
\begin{equation}
t\simeq \frac{R}{4\gamma_2^2c}.
\end{equation}

Using equations (12) and (14), the Lorentz factor of the shocked
medium (i.e., region 2) just behind the forward shock is found to
evolve with time as
\begin{equation}
\gamma_2=\left(\frac{17E_0}{1024\pi n_1m_pc^5t^3}\right)^{1/8}.
\end{equation}
From equations (10)-(13), we have the similarity variable at the
location of the contact discontinuity,
\begin{equation}
\chi=\left(\frac{4L_wt}{17E_0}\right)^{-12/17}
=3.1\left(\frac{L_{w,47}t}{E_{52}}\right)^{-12/17},
\end{equation}
and the Lorentz factor of region 3,
\begin{eqnarray}
\gamma_3 & = & \left[\frac{(4L_w)^{12/17}(17E_0)^{5/17}}{1024\pi
n_1m_pc^5t^{39/17}}\right]^{1/8}\nonumber \\ & = &
5.3L_{w,47}^{3/34}E_{52}^{5/136}n_1^{-1/8}t^{-39/136}
\end{eqnarray}
where $L_{w,47}=L_w/10^{47}\,{\rm erg}\,{\rm s}^{-1}$,
$E_{52}=E_0/10^{52}\,{\rm ergs}$, and $n_1$ and $t$ are in units
of $1\,{\rm cm}^{-3}$ and 1 day respectively.

Letting $\chi=1$, we define a critical time
\begin{equation}
t_{\rm cr}=4.9E_{52}L_{w,47}^{-1}\,{\rm days}.
\end{equation}
At this time, the injection energy to the fireball significantly
exceeds its initial energy. For $t<t_{\rm cr}$, the similarity
variable $\chi>1$. It should be emphasized that the dynamics
denoted by equation (17) is simply calculated by equating the
pressures of the two-sided shocked fluids at the contact
discontinuity. In this derivation, we have neglected any work done
on region 2 by region 3 because the pressure of region 3 is much
less than $P'_2(R)$ at $t<t_{\rm cr}$.

Once the observer's time exceeds $t_{\rm cr}$, the similarity
variable $\chi=1$. At this stage, the total kinetic energy of
region 2 is approximated as $E_{{\rm kin},2}=(\gamma_2^2-1)M_{\rm
sw}c^2$, where $M_{\rm sw}=(4\pi/3)R^3n_1m_p$ is the swept-up
medium mass. Energy conservation requires that any increase of
kinetic energy of region 2 should be equal to work done by region
3,
\begin{equation}
dE_{{\rm kin},2}=\gamma_3P'_3dV'_3,
\end{equation}
where $dV'_3=4\pi R^2dR'=4\pi R^2(dR/\gamma_3)$ is the volume
change of region 3 in its own rest frame. Since regions 2 and 3
should keep velocity equality along the contact discontinuity
(viz., $\gamma_2=\gamma_3$), we rewrite equation (19) as
\begin{equation}
\frac{d\gamma_2}{dR}=\frac{4\pi
R^2[P'_3-(\gamma_2^2-1)n_1m_pc^2]}{2\gamma_2M_{\rm sw}c^2}.
\end{equation}
Considering equation (5) and the shock radius $R\simeq
4\gamma_2^2ct$, the solution of equation (20) becomes
\begin{equation}
\gamma_2=\gamma_3=\left(\frac{L_w}{128\pi
n_1m_pc^5t^2}\right)^{1/8},
\end{equation}
where the dependence of $\gamma_2$ on $t$ is consistent with the
one derived for a pure electromagnetic energy injection by Dai \&
Lu (1998).

We next discuss the dynamics of a relativistic wind bubble: In the
case of $t_{\rm cr}>T_{{\rm M},0}$ (viz.,
$E_{52}>0.46I_{45}P_{0,{\rm ms}}^{-2}$), the wind bubble should
evolve based on equations (15) and (17); for $t_{\rm cr}<T_{{\rm
M},0}$ (viz., $E_{52}<0.46I_{45}P_{0,{\rm ms}}^{-2}$), however,
the Lorentz factors of the wind bubble decay initially as
$\gamma_2\propto t^{-3/8}$ and $\gamma_3\propto t^{-39/136}$ at
$t<t_{\rm cr}$ (stage I), subsequently as
$\gamma_2=\gamma_3\propto t^{-1/4}$ at $t\in (t_{\rm cr},T_{{\rm
M},0})$ (stage II), and finally again as $\gamma_2\propto
t^{-3/8}$ at $t>T_{{\rm M},0}$ (stage III). It should be pointed
out that this discussion assumes a negligible radiative loss of
region 3. However, no matter whether region 3 is at stage I or II,
and once it enters the fast cooling regime at some time, its
pressure will begin to become much smaller than that of region 2
and then the Lorentz factor of region 2 will decay as equation
(15).

\section{Synchrotron Radiation and Light Curves}

In this section we discuss light curves of the emission from a
relativistic wind bubble, assuming $t_{\rm cr}<T_{{\rm M},0}$. The
dynamics above determines the bulk Lorentz factor and the thermal
Lorentz factors of the accelerated electrons of each region as
function of time. We consider synchrotron radiation from each
region at different stages. According to the standard afterglow
model (M\'esz\'aros \& Rees 1997b; Sari, Piran \& Narayan 1998),
the spectrum consists of four power-law segments separated by
three break frequencies: the self-absorption frequency $\nu_a$,
the characteristic frequency $\nu_m$, and the cooling frequency
$\nu_c$, with the peak flux $F_{\nu,{\rm max}}$. To calculate
them, we assume that for region 3 the electron and magnetic field
energy densities are fractions $\epsilon_e$ and $\epsilon_B$ of
the total energy density behind the reverse shock (where
$\epsilon_e+\epsilon_B=1$), but for region 2, fractions $\xi_e$
and $\xi_B$ of the total energy density behind the forward shock
(where $\xi_e+\xi_B <1$). One may expect that
$\epsilon_e\neq\xi_e$ and/or $\epsilon_B\neq\xi_B$, as suggested
in some recent studies (Coburn \& Boggs 2003; Zhang, Kobayashi \&
M\'esz\'aros 2003; Kumar \& Panaitescu 2003). In addition, we
assume that the spectral index of the electron energy distribution
is $p$ for region 3 and $q$ for region 2.

At stage I ($t<t_{\rm cr}$), the break frequencies and peak flux
of region 3 are derived as
\begin{eqnarray}
\nu_{m,3}^{\rm I} & = & 6.8\times
10^{12}g_p^2\epsilon_e^2\epsilon_{B,-1}^{1/2}
\gamma_{w,4}^2\nonumber \\ & & \times L_{w,47}^{5/34}
E_{52}^{-5/34}n_1^{1/2}t^{5/34}\,{\rm
Hz},\\
\nu_{c,3}^{\rm I} & = & 3.0\times 10^{14}\epsilon_{B,-1}^{-3/2}
   L_{w,47}^{-27/34}\nonumber \\ & & \times E_{52}^{5/17}
   n_1^{-1}t^{-22/17}\,{\rm Hz},\\
F_{\nu,{\rm max},3}^{\rm I} & = &
88\epsilon_{B,-1}^{1/2}\gamma_{w,4}^{-1}L_{w,47}^{45/34}
   E_{52}^{-5/68}\nonumber \\ & & \times n_1^{1/4}
   D_{L,28}^{-2}t^{39/68}\,{\rm mJy},
\end{eqnarray}
where $g_p=(p-2)/(p-1)$, $\epsilon_{B,-1}=\epsilon_B/0.1$,
$\gamma_{w,4}=\gamma_w/10^4$, and $D_{L,28}$ is the luminosity
distance to the source in units of $10^{28}$ cm. We here
considered only the characteristic frequency and the cooling
frequency for the optical to X-ray emission discussed in this
paper. In our derivation, we used equations (8), (10) and (17) to
obtain the energy density of region 3,
$e'_3=4n_1m_pc^2\gamma_2^2\chi^{-17/12}=4n_1m_pc^2\gamma_3^2
\chi^{-5/12}=0.11L_{w,47}^{8/17}E_{52}^{-15/68}n_1^{3/4}t^{-19/68}
\,{\rm erg}\,{\rm cm}^{-3}$, and the magnetic field strength,
$B'_3=(8\pi\epsilon_Be'_3)^{1/2}=0.52\epsilon_{B,-1}^{1/2}
L_{w,47}^{4/17}E_{52}^{-15/136}n_1^{3/8}t^{-19/136}$ G. We define
a cooling time $t_{0,3}^{\rm I}$ through $\nu_{m,3}^{\rm
I}=\nu_{c,3}^{\rm I}$ as
\begin{eqnarray}
t_{0,3}^{\rm I} & = &
14(g_p\epsilon_e\epsilon_{B,-1}\gamma_{w,4})^{-68/49}\nonumber \\
& & \times L_{w,47}^{-32/49} E_{52}^{15/49}n_1^{-51/49}\,{\rm
days}.
\end{eqnarray}
From equations (22) and (23), we find that region 3 is in the
slow-cooling regime for $t<t_{0,3}^{\rm I}$ but in the
fast-cooling regime for $t>t_{0,3}^{\rm I}$, in contrast to the
standard afterglow model (Sari et al. 1998). A larger value of
$\gamma_w$ would imply fast cooling in region 3 earlier on, in
which case region 2 could still stay at stage I. For region 2, the
break frequencies and peak flux are given by
\begin{eqnarray}
\nu_{m,2}^{\rm I} & = & 1.6\times
10^{13}g_q^2\xi_{e,-1}^2\xi_{B,-1}^{1/2}
E_{52}^{1/2}t^{-3/2}\,{\rm Hz},\\
\nu_{c,2}^{\rm I} & = & 8.5\times 10^{13}\xi_{B,-1}^{-3/2}
         E_{52}^{-1/2}n_1^{-1}t^{-1/2}\,{\rm Hz},\\
F_{\nu,{\rm max},2}^{\rm I} & = & 35\xi_{B,-1}^{1/2}
         E_{52}n_1^{1/2}D_{L,28}^{-2}\,{\rm mJy},
\end{eqnarray}
where $g_q=(q-2)/(q-1)$, $\xi_{e,-1}=\xi_e/0.1$, and
$\xi_{B,-1}=\xi_B/0.1$ (Sari et al. 1998).

At stage II ($t_{\rm cr}<t<T_{{\rm M},0}$), since $\gamma_2\propto
t^{-1/4}$, we obtain the break frequencies and the peak flux for
region 3,
\begin{equation}
\nu_{m,3}^{\rm II}\propto t^0,\,\, \nu_{c,3}^{\rm II}\propto
t^{-1},\,\, F_{\nu,{\rm max},3}^{\rm II}\propto t^{1/2},
\end{equation}
and for region 2,
\begin{equation}
\nu_{m,2}^{\rm II}\propto t^{-1},\,\, \nu_{c,2}^{\rm II}\propto
t^{-1},\,\, F_{\nu,{\rm max},2}^{\rm II}\propto t^1.
\end{equation}
We define another cooling time $t_{0,3}^{\rm II}$ at which
$\nu_{m,3}^{\rm II}= \nu_{c,3}^{\rm II}$ as follows
\begin{equation}
t_{0,3}^{\rm II}=22(g_p\epsilon_e\epsilon_{B,-1}\gamma_{w,4})^{-2}
L_{w,47}^{-1/2}n_1^{-3/2}\,{\rm days}.
\end{equation}
Equation (29) is valid only for $t<\min (t_{0,3}^{\rm II}, T_{{\rm
M},0})$. Otherwise, once region 3 enters the fast-cooling regime
or the wind luminosity weakens obviously, its pressure becomes
insignificant as compared to that of region 2 and thus evolution
of the wind bubble comes back to $\gamma_2\propto t^{-3/8}$.

We assume $t_{0,3}^{\rm II}<T_{{\rm M},0}$. If $t_{\rm
cr}<t<t_{0,3}^{\rm II}$, then region 3 is adiabatic and its
emission spectrum is determined by equation (29). If $t_{0,3}^{\rm
II}<t<T_{{\rm M},0}$, region 3 becomes fast cooling, and the break
frequencies and the peak flux for region 2 evolve as
$\nu_{m,2}^{\rm II}\propto t^{-3/2}$, $\nu_{c,2}^{\rm II}\propto
t^{-1/2}$ and $F_{\nu,{\rm max},2}^{\rm II}\propto t^0$. At stage
III, the emission flux of region 3: $F_{\nu,3}^{\rm III}\propto
\nu^{-\beta}t^{-(2+\beta)}$, where $\beta$ is the spectral index
at $t_{0,3}^{\rm II}$ (Kumar \& Panaitescu 2000).

On the other hand, $t_{0,3}^{\rm II}>T_{{\rm M},0}$ is assumed. If
$t_{\rm cr}<t<T_{{\rm M},0}$, the spectrum and light curves for
regions 3 and 2 are obtained by using equations (29) and (30),
respectively. But if $t>T_{{\rm M},0}$ (stage III), because of
adiabatic expansion of region 3, the break frequencies and the
peak flux decay as $\nu_{m,3}^{\rm III}\propto t^{-73/48}$,
$\nu_{{\rm cut},3}^{\rm III}\propto t^{-73/48}$ and $F_{\nu,{\rm
max},3}^{\rm III}\propto t^{-47/48}$ (Sari \& Piran 1999).

According to the derived scaling laws of the break frequencies and
peak flux with time at three stages, we can obtain the light curve
indices for different frequency bands in the case of $t_{\rm
cr}<T_{{\rm M},0}$ (see Table 1). As an example, Figure 1 presents
R-band light curves for typical values of the model parameters:
$I_{45}=3$, $P_{0,{\rm ms}}=1$, $\gamma_{w,4}=1$, $L_{w,47}=1$,
$p=q=2.5$, $\epsilon_e=0.9$, $\epsilon_B=\xi_e=\xi_B=0.1$,
$E_{52}=1$, $n_1=1\,{\rm cm}^{-3}$, and $D_{L,28}=1$. We can see
that the emission flux from region 2 decays rapidly at time
$<t_{\rm cr}$, subsequently fades more slowly at time $\in(t_{\rm
cr}, T_{{\rm M},0})$, and finally declines based on the initial
evolution law (Dai \& Lu 1998). More importantly, the emission
from region 3 dominates the afterglow emission, which leads to a
bump in the afterglow light curve.

\begin{table*}[ht!]
\begin{tabular}{|c|c|c|c|c|c|c|c|}
\hline & & stage I & \multicolumn{3}{|c|}{stage II ($t_{\rm
cr}<t<T_{{\rm M},0}$)}
& \multicolumn{2}{|c|}{stage III} \\
region & frequency & slow cooling &  $t<t_{0,3}^{\rm II}<T_{{\rm
M},0}$ &  $t_{0,3}^{\rm II}<t<T_{{\rm M},0}$ & $t_{0,3}^{\rm
II}>T_{{\rm M},0}$ &   $t_{0,3}^{\rm II}<T_{{\rm M},0}$ &
$t_{0,3}^{\rm II}>T_{{\rm M},0}$
\\\hline\hline
3... & $\nu<\nu_p$ & $-{107\over 204}$ & $-{1\over 2}$ &
$-{17\over 12}$ & $-{1\over 2}$ & $5\over 3$ & $17\over 36$
\\
& $\nu_p<\nu<\nu_0$ & $-{5p+34 \over 68}$ & $-{1\over 2}$ &
${1\over 4}$ & $-{1\over 2}$ & ${p+3\over 2}$ & $73p+21\over 96$
\\
& $\nu>\nu_0$ & $-{5(p-2)\over 68}$ & $0$ & $-{p-2\over 4}$ & $0$
& $p+4\over 2$ & $p+3\over 2$
\\\hline
2... & $\nu<\nu_p$ & $-{1\over 2}$ & $-{4\over 3}$ & $-{1\over 2}$
& $-{4\over 3}$ & $-{1\over 2}$ & $-{1\over 2}$
\\
& $\nu_p<\nu<\nu_0$ & $3(q-1)\over 4$ & $q-3\over 2$ &
$3(q-1)\over 4$ & $q-3\over 2$ & $3(q-1)\over 4$ & $3(q-1)\over 4$
\\
& $\nu>\nu_0$ & $3q-2\over 4$ & $q-2\over 2$ & $3q-2\over 4$ &
$q-2\over 2$ & $3q-2\over 4$ & $3q-2\over 4$
\\\hline\hline
\end{tabular}
\par
\label{t:afterglow} \caption{The light curve index $\alpha$ as
function of $p$ or $q$ ($F_\nu\propto t^{-\alpha}$). Definition:
$\nu_p=\min(\nu_m, \nu_c)$ and $\nu_0=\max(\nu_m, \nu_c)$}
\end{table*}

\section{Discussion and Conclusions}

Based on the successful models of the Crab Nebula, we discuss the
dynamics of a relativistic wind bubble and its emission
signatures. Such a wind bubble is naturally expected when an
ultrarelativistic e$^+$e$^-$-pair wind from a highly magnetized,
rapidly rotating object at the center of an afterglow interacts
with an outward-expanding fireball, regardless of whether this
object is a millisecond magnetar or a Kerr black hole. We find
that when the injection energy significantly exceeds the initial
energy of the fireball, the bulk Lorentz factor of the wind bubble
declines more slowly than before. In addition, the reverse-shock
emission could dominate the afterglow emission, which leads to a
bump in afterglow light curves. In this paper, we discuss the case
of $t_{\rm cr}<T_{{\rm M},0}$. However, even if $t_{\rm
cr}>T_{{\rm M},0}$, a bump in Figure 1 still appears for some
parameter space as we move the late-time light curve of region 3
to early times.

Bump features have been detected in some events (e.g., GRBs
970508, 000301C, 021004, and 030329). To interpret these features,
some other models invoked the microlensing event (Garnavich, Loeb
\& Stanek 2000), density-jump medium (Dai \& Lu 2002; Lazzati et
al. 2002; Dai \& Wu 2003), pure Poynting-flux injection (Dai \& Lu
1998; Zhang \& M\'esz\'aros 2001), baryon-dominated injection
(Rees \& M\'esz\'aros 1998; Sari \& M\'esz\'aros 2000; Zhang \&
M\'esz\'aros 2002; Granot, Nakar \& Piran 2003), and two-component
jet (Berger et al. 2003; Huang et al. 2003). The magnetic field in
the reversely-shocked region of the wind bubble seems to consist
of two components: a toroidal field and a random field. The latter
may be naturally generated by the relativistic two-stream
instability (Medvedev \& Loeb 1999). If the toroidal field
dominates over the random component, one would expect high
polarization of the bump emission, which could be used to
distinguish between our wind-bubble model and other explanations.
It is well known that the degree of linear polarization of
synchrotron radiation from the reversely-shocked wind is about
$\Pi_{\rm syn}=(p+1)/(p+7/3)$ for a large-scale toroidal magnetic
field. Thus, the polarization of an afterglow in our model could
be as high as $\Pi_{\rm syn}\sim 70\%$ for $p\sim 2$ when the
reverse shock emission dominates the afterglow emission {\em
during an obvious bump}. Even if the total afterglow flux is
dominated by the forward shock emission, and the reversely-shocked
wind provides only a small fraction of the total flux, $\zeta$,
then the shocked pair wind could still dominate the polarized flux
with linear polarization of $\Pi\sim \zeta\Pi_{\rm syn}=
7\%(\zeta/0.1)(\Pi_{\rm syn}/0.7)$. This value is larger than the
currently observed data of GRBs 021004 and 030329 ($\sim 2-3\%$,
Rol et al. 2003; Lazzati et al. 2003; Greiner et al. 2003). If, on
the other hand, the random-field strength exceeds the toroidal
component, the calculated degree of linear polarization is
significantly less than that estimated above.

We have considered a spherical wind bubble in our paper. However,
there is evidence that GRBs are collimated into narrow jets, whose
kinetic energy is clustered at $E_{\rm jet}\sim 10^{50}-10^{51}$
ergs (Frail et al. 2001; Panaitescu \& Kumar 2002; Berger,
Kulkarni \& Frail 2003; Bloom, Frail \& Kulkarni 2003). On the
other hand, relativistic winds from millisecond magnetars are
roughly isotropic and their total energy is $\sim 10^{53}$ ergs.
Beaming correction gives an injection energy of $\sim 10^{51}$
ergs within the initial solid angle of a jet. This energy is of
the same order as $E_{\rm jet}$. Actually, the jet can get more
energy from its central magnetar due to sideways expansion. This
would favor our model.

\acknowledgments The author thanks the referee and B. Zhang for
valuable comments and suggestions, and Y. F. Huang, X. Y. Wang, D.
M. Wei and X. F. Wu for useful discussions. This work was
supported by the National Natural Science Foundation of China
(grants 10233010 and 10221001) and the National 973 Project
(NKBRSF G19990754).

\clearpage
\begin{figure}
\begin{picture}(100,250)
\put(0,0){\includegraphics{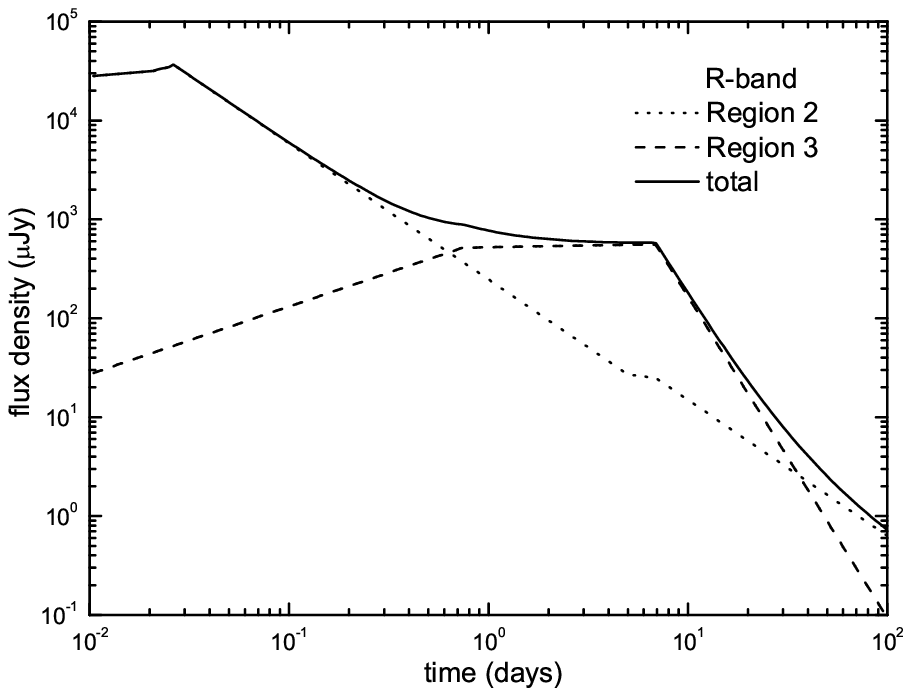}}
\end{picture}
\caption {R-band light curves of the emissions from regions 2
(dotted line) and 3 (dashed line). The solid line corresponds to
the total flux.}
\end{figure}

\end{document}